\documentclass[12pt]{elsart}
\usepackage{amssymb}
\usepackage{graphicx}
\begin{document}
\begin{frontmatter}

\title{Development of Thick-foil and Fine-pitch GEMs with a Laser
Etching Technique}

\author[label1,label2]{T. Tamagawa}
\author[label1,label2]{A. Hayato}
\author[label1,label2]{F. Asami}
\author[label1,label2]{K. Abe}
\author[label1,label3]{S. Iwamoto}
\author[label1,label2]{S. Nakamura}
\author[label4]{A. Harayama}
\author[label1,label2]{T. Iwahashi}
\author[label1,label2]{S. Konami}
\author[label5]{H. Hamagaki}
\author[label5]{Y. L. Yamaguchi}
\author[label6]{H. Tawara}
\author[label7,label1]{K. Makishima}

\address[label1]{RIKEN, 2-1 Hirosawa, Wako, Saitama 351-0198, Japan}
\address[label2]{Department of Physics, Tokyo University of Science, 1-3
 Kagurazaka, Shinjyuku-ku, Tokyo 162-8601, Japan}
\address[label3]{Department of Physics, Tokai University, 1117
 Kitakaname, Hiratuka, Kanagawa 259-1292, Japan}
\address[label4]{Department of Physics, Saitama University, 255
Shimo-Okubo, Sakura-ku, Saitama City, Saitama 338-8570, Japan}
\address[label5]{Center for Nuclear Study (CNS), University of Tokyo,
 2-1 Hirosawa, Wako, Saitama 351-0198, Japan}
\address[label6]{KEK, 1-1 Oho, Tsukuba, Ibaraki 305-0801, Japan}
\address[label7]{Department of Physics, University of Tokyo, 7-3-1
 Hongo, Bunkyo-ku, Tokyo 113-0033, Japan}
 
 \begin{abstract}
  We have produced thick-foil and fine-pitch gas electron multipliers
  (GEMs) using a laser etching technique. To improve production yield we
  have employed a new material, Liquid Crystal Polymer, instead of
  polyimide as an insulator layer. The effective gain of the thick-foil
  GEM with a hole pitch of 140 $\mu$m, a hole diameter of 70 $\mu$m, and
  a thickness of 100 $\mu$m reached a value of 10$^4$ at an applied
  voltage of 720~V. The measured effective gain of the thick-foil and
  fine-pitch GEM (80 $\mu$m pitch, 40 $\mu$m diameter, and 100 $\mu$m
  thick) was similar to that of the thick-foil GEM. The gain stability
  was measured for the thick-foil and fine-pitch GEM, showing no
  significant increase or decrease as a function of elapsed time from
  applying the high voltage. The gain stability over 3 h of operation
  was about 0.5\%. Gain mapping across the GEM showed a good uniformity
  with a standard deviation of about 4\%. The distribution of hole
  diameters across the GEM was homogeneous with a standard deviation of
  about 3\%. There was no clear correlation between the gain and hole
  diameter maps.
 \end{abstract}


\end{frontmatter}

 \section{Introduction}

Gas electron multipliers (GEMs) are one of the more recently developed
micro-pattern gas detectors \cite{sauli1997}. Dense through-holes are
drilled in an insulator substrate sandwiched by thin copper foils. By
applying high voltage to the copper electrodes in an appropriate gas, we
are able to use the GEM as an electron multiplier in which each hole
works as an individual proportional counter. This device is used for
many applications such as time projection chambers \cite{kobayashi2007},
photon detectors \cite{meinschad2004}, X-ray imagers \cite{sauli2003},
two phase xenon detectors for dark matter search \cite{balau2009}, etc.

GEMs are generally produced by using a chemical etching technique
\cite{cern-gdd}. As an alternative, we have successfully developed GEMs
using a laser etching technique \cite{tamagawa2006}, which has many
advantages.  Cylindrical holes are easily formed with the laser method,
while double-conical holes are formed with the chemical etching
technique \cite{cern-gdd}. The capability to drill cylindrical holes
helps to form finer pitch holes on a thicker substrate. With the laser
etching technique we have achieved 50 $\mu$m pitch holes on a 50 $\mu$m
thick foil \cite{tamagawa2006}.

The primary purpose of our GEM development is to construct a
photoelectric X-ray polarimeter for space missions, particularly for
astrophysics. Recently, several photoelectric polarimeters have been
proposed \cite{costa2001,kevin2007,poet2008}. For the polarimeters, the
GEM is a key device to achieve the mission goals. The details of
measuring the polarization of X-rays are described in
Refs.\cite{bellazzini2008,kevin2007}. The key to high sensitivity is to
measure the track of the photoelectron in the gas (typically a few 100
$\mu$m) with sufficient accuracy. According to our simulations the GEM
should have a good spatial resolution of 50--100~$\mu$m.

The most significant risk to operating a GEM in space are
discharges. The electron multiplication factor (gain) of a ``standard''
GEM with a pitch of 140 $\mu$m, a hole diameter of 70 $\mu$m, and a
thickness of 50 $\mu$m, is a few 1000 in a gas mixture of 70\% argon and
30\% carbon dioxide by volume (Ar-70/CO$_2$-30). Although a gain of 1000
is sufficient when a very low-noise readout ASIC is combined with the
GEM, we require that the GEM works without any discharge up to a gain of
around 10$^4$ for safety in space. It is known that the gain of a double
or triple GEM configuration can easily reach a value of 10$^4$. The
stack of GEMs, however, diffuses electrons in the transfer section.

Recently, the so-called ``thick GEMs (THGEMs)'', in which through-holes
are mechanically drilled on a glass epoxy substrate, have been
fabricated to achieve robustness against electrical discharge
\cite{breskin2009}. THGEMs are already used for some applications such
as ultraviolet photon counting detectors. Mechanical drilling is a
well-developed technique, and the glass epoxy substrate may be easier to
treat than the flexible polymer used in ordinary GEMs. It is, however,
impossible to mechanically drill fine-pitch through-holes, e.g. less
than 100 $\mu$m pitch, on the glass epoxy substrate. Therefore, the
THGEMs are not suitable for this application.

 \section{Production of Thick-foil and Fine-pitch GEMs}

Table \ref{tab:gem_list} presents the list of the GEMs that were used in
this study.\footnote{GEM foils produced by SciEnergy Co., Ltd.
(info@scienergy.jp)} The production process was the same as those shown
in our previous paper \cite{tamagawa2006}. The active area of all GEMs
that we tested was 30$\times$30~mm$^2$, although SciEnergy can provide
up to 230$\times$300 mm$^2$ GEMs. First, we fabricated the GEMs referred
to as ``RIKEN-140-PI'' which had the same hole pitch (140 $\mu$m),
diameter (70 $\mu$m), and thickness (50 $\mu$m) as the so-called
``standard'' CERN GEMs. The only difference between the RIKEN-140-PI and
the CERN standard GEM was the method for drilling holes through the
substrate. We have employed a carbon dioxide laser
($\lambda=$10.6~$\mu$m), whereas CERN uses a wet etching technique.

Second, we produced the GEMs by using a new material, Liquid Crystal
Polymer (LCP)\footnote{The trade name of the LCP substrate we used is
the ESPANEX$^{\textregistered}$ L-Series manufactured by Nippon Steel
Chemical Co., Ltd.}, instead of using polyimide (PI) as the insulating
layer. The first LCP GEM we produced, referred to as ``RIKEN-140-LCP'',
has the same hole pitch, diameter, and thickness as the
RIKEN-140-PI. The mechanical, thermal, and hygroscopic properties of LCP
are summarized in Table \ref{tab:lcp_parameters}. LCP gives lower
expansion rate in fabrication as it absorbs less water than PI. This is
an advantage for etching micro-pattern foils with a good accuracy. The
low moisture absorption rate also leads to less out-gassing in
operations. Although the melting temperature of LCP is slightly lower
than that of PI, this is not a disadvantage for normal operation.

\begin{figure}[h]
  \begin{center}
   \includegraphics[width=0.55\textwidth]{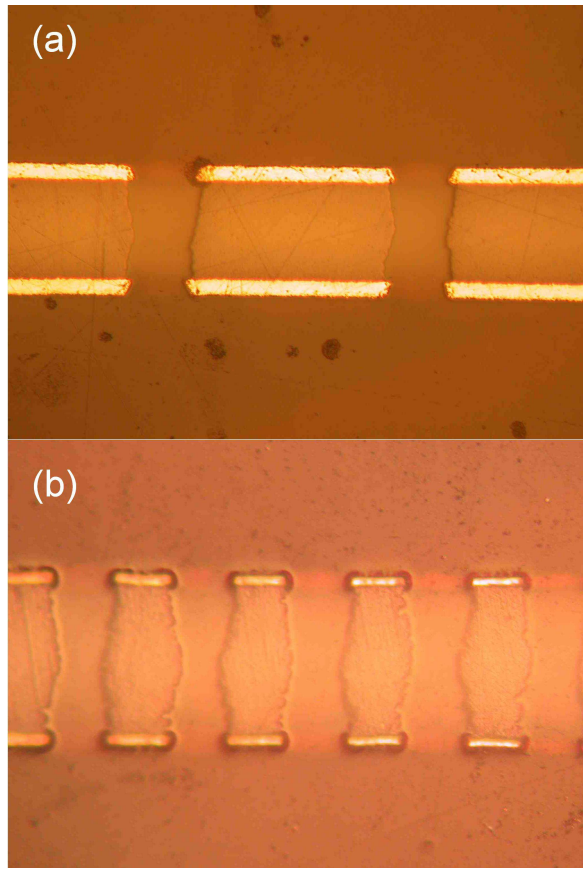}
  \end{center}
  \caption{Cross-section of (a) RIKEN-140T-LCP and (b) RIKEN-80T-LCP
  obtained with a metallographic microscope.}
  \label{fig:cross_sections}
 \end{figure}

Third, we fabricated GEMs which had 140 $\mu$m pitch, 70 $\mu$m
diameter, and 100 $\mu$m thick LCP insulator, referred to as
``RIKEN-140T-LCP''. According to a simulation to calculate the electric
field inside the holes, we expected that a thicker foil would yield more
gain than a 50 $\mu$m thick GEM at the same applied high voltage. This
was the motivation for producing the thick-foil GEMs. When we employed
the PI substrate, we produced very few of 100 $\mu$m thick GEMs. After
we changed the substrate from PI to LCP, the production yield of the
thick-foil GEMs was improved and reached almost 100\%. A cross-section
of the GEM is shown in Figure \ref{fig:cross_sections}a.

Finally, we fabricated fine-pitch holes on a 100 $\mu$m thick LCP
substrate. The GEMs have 80 $\mu$m pitch and 40 $\mu$m diameter,
referred to as ``RIKEN-80T-LCP''. A cross-section of the GEM is shown in
Figure \ref{fig:cross_sections}b. The fine-pitch and thick-foil GEM was
expected to give a high gain without any discharge while keeping the
high spatial resolution.

 \section{Gain Measurements and Results}

  \subsection{Test setup}
  \label{sect:test_setup}

\begin{figure}[h]
   \begin{center}
    \includegraphics[width=1.0\textwidth]{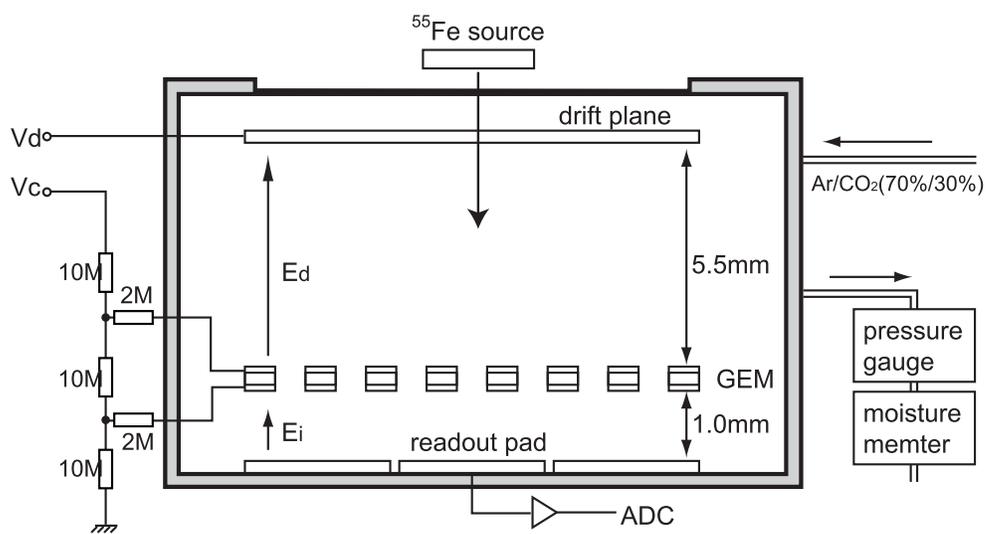}
   \end{center}
   \caption{A schematic view of the GEM test setup.}
   \label{fig:exp_setup}
  \end{figure}

  Figure \ref{fig:exp_setup} shows a schematic view of the GEM test
  setup used in this study. The setup consisted of a drift plane, a GEM
  foil, and 3$\times$3 readout pads each with an area of
  9$\times$9~mm$^2$. The spacing between two adjacent pads was
  1~mm. Only the central pad was read out and the others were connected
  to ground. The drift plane was a 15 $\mu$m thick aluminum foil with an
  active area of 30$\times$30 mm$^2$. The drift plane, GEMs, and readout
  pads were placed in a chamber which was then filled with gas. The
  vertical spacing of the target region, which was the spacing between
  the drift plane and GEM, was 5.5~mm, and the induction region between
  GEM and the readout pad was 1.0~mm. A high voltage was supplied via a
  chain of 10 M$\Omega$ resistors, and to minimize the risk of electric
  surges, a 2 M$\Omega$ protection resistor was added in series with
  each GEM electrode. The electric field in the drift region was
  E$_d$=2.5 kV cm$^{-1}$, and in the induction region, E$_i$=4--5 kV
  cm$^{-1}$.
  
  During the test, Ar-70/CO$_2$-30 gas was made to flow through the
  system. The primary reason we selected this gas mixture was to easily
  compare our results to other experiments; many GEM studies have been
  done with this gas mixture. In this study, we did not add any
  additional gases to the mixture to prevent discharge, aging effects,
  etc. As shown in Figure~\ref{fig:exp_setup}, we equipped the chamber
  with a pressure gauge and a moisture meter as well as a
  thermometer. The experiments were carried out in an air-conditioned
  room at a temperature of around 20~$^{\circ}$C. The moisture in the
  gas was kept at 50~ppm or less. The pressure and temperature of the
  gas were recorded automatically every 10 min or faster.
    
  Charge signals from the readout pads were fed into an AmpTek A225,
  which consisted of a charge sensitive preamplifier and shaper
  module. The amplified and shaped voltage signals were fed into the
  custom-made main amplifier. This module had a discriminator and gate
  generator for data acquisition system control. The amplified signals
  were fed into a VME peak-hold ADC (Clear Pulse 1113A) controlled by a
  PC. To make a calibration curve between the amount of input charge and
  ADC channel, a well-defined rectangular wave from a research pulser
  (ORTEC model 448) was fed into the preamplifier through a 2~pF
  capacitor.

  The effective gain ($G_{\rm eff}$) is given by;

  \begin{equation}
   G_{\rm eff}=Const\times\frac{S_{\rm mean}}{q_{\rm e} \; n_{\rm e}}
  \end{equation}
  where $S_{\rm mean}$ is the ADC peak value of incident monochromatic
  X-rays, $q_{\rm e}$ is the electron charge
  (1.602$\times$10$^{-19}$~C), and $n_{\rm e}$ is the number of
  electron-ion pairs created by an X-ray photon. A typical value of
  $n_{\rm e}$ is 212 for a 5.9~keV X-ray photon from the radioactive
  $^{55}$Fe source in the Ar-70/CO$_2$-30 gas mixture
  \cite{seed_electron}. Typical energy resolution was 18\% (FWHM) at
  5.9~keV. The relative systematic uncertainty of the measurement was
  about 1\% of the gain value as far as we used the readout system
  described above. When we used other readout systems to estimate the
  absolute systematic uncertainty of the gain value, the maximum shift
  was 6\%.

  \subsection{Gain curves}

  \subsubsection{Polyimide and LCP GEMs}

\begin{figure}[h]
   \begin{center}
    \includegraphics[width=1.0\textwidth]{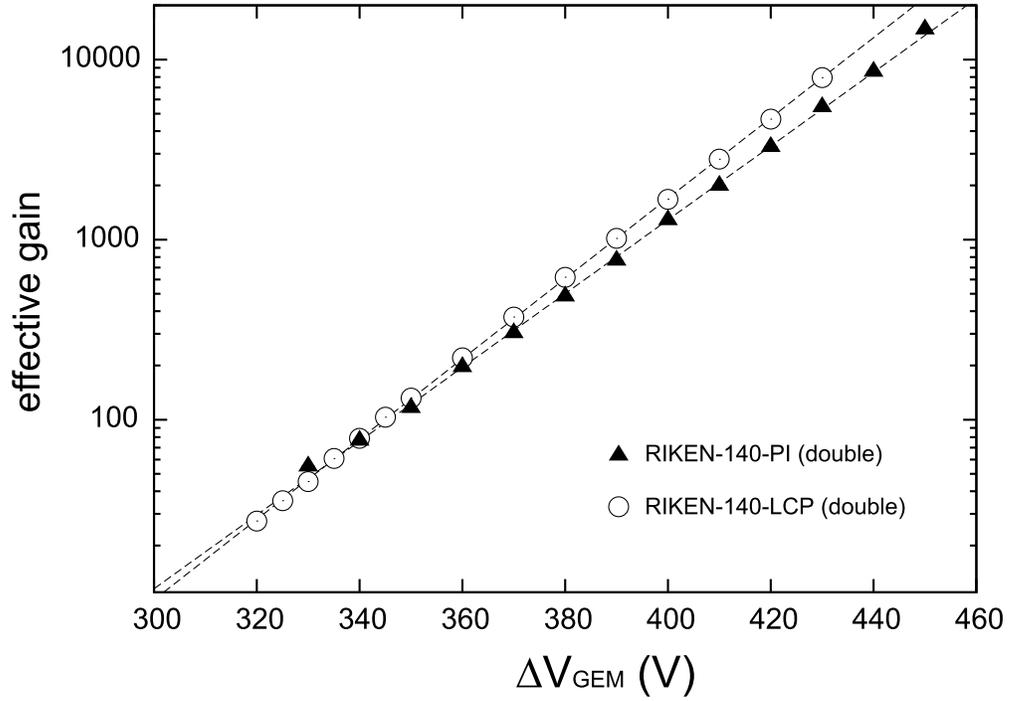}
   \end{center}
   \caption{Effective gain of double RIKEN-140-LCPs and double
   RIKEN-140-PIs as a function of the applied voltage to each GEM.}
   \label{fig:gain_curve_pi_lcp}
  \end{figure}

  Figure~\ref{fig:gain_curve_pi_lcp} shows the gain curves of double
  RIKEN-140-PI and double RIKEN-140-LCP GEM stacks. Both gain curves
  show that an effective gain of around 10$^4$ was achieved at an
  applied high voltage of 430~V per GEM. No difference in gain was
  observed between the PI and LCP GEMs, indicating that the difference
  in the insulator materials is not essential for the multiplication
  mechanism of electrons in the holes. The gain curve of the CERN
  standard GEM was almost the same as the RIKEN-140-PI and RIKEN-140-LCP
  GEMs as shown in Figure 6 of our previous paper
  \cite{tamagawa2006}. Since there was no significant difference between
  these three GEMs, we concluded that using a LCP substrate together
  with laser-drilled holes is an ideal method to produce GEMs,
  equivalent to other methods such as laser or wet etching with PI
  substrates.

   \subsubsection{Thick-foil GEMs}
   
\begin{figure}[h]
   \begin{center}
    \includegraphics[width=1.0\textwidth]{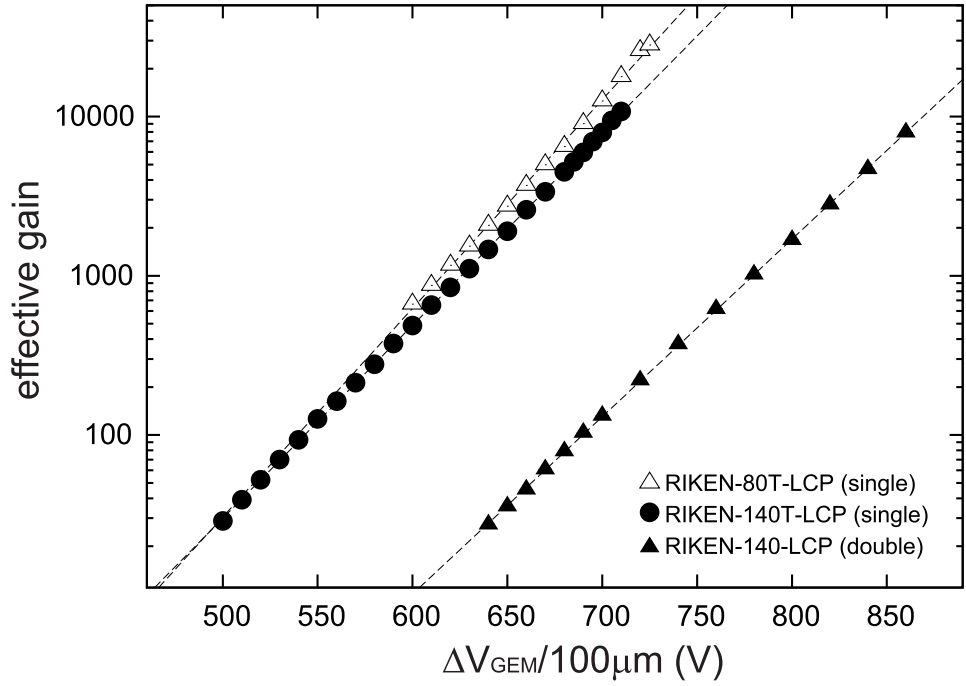}
   \end{center}
    \caption{Effective gain of single RIKEN-140T-LCP, single
    RIKEN-80T-LCP, and double RIKEN-140-LCP GEMs.}
    \label{fig:gain_curve_thickfoil}
  \end{figure}

   Figure \ref{fig:gain_curve_thickfoil} shows the gain curve of a
   single RIKEN-140T-LCP GEM. An effective gain of 10$^4$ was achieved
   at an applied voltage of 700~V per GEM. For comparison, a double
   RIKEN-140-LCP stack, in which the two GEMs were vertically spaced by
   1~mm with an electric field between them of about 3.5~kV cm$^{-1}$,
   is superposed on the same figure. The effective gain of the
   thick-foil RIKEN-140T-LCP GEM was about 60 times higher than that of
   the double RIKEN-140-LCP GEM stack, when we operated the double GEM
   at the same combined voltage as the thick-foil GEM. Obtaining an
   appropriate gain at lower applied voltage decreases the risk of
   discharges. We can also avoid electron diffusion in the transfer
   region between two GEMs.
   
   \subsubsection{Thick-foil and fine-pitch GEMs}

   A gain curve of a single RIKEN-80T-LCP GEM is shown in
   Figure~\ref{fig:gain_curve_thickfoil}. The maximum gain obtained in
   Ar-70/CO$_2$-30 gas, without any micro-discharge was 3$\times$10$^4$
   at a high voltage of 725~V.  Although the hole size of RIKEN-80T-LCP
   was smaller than that of RIKEN-140T-LCP, the gain curves were almost
   identical. This similarity was predicted from our previous study in
   which we demonstrated independence of the gain on the hole size, for
   50 $\mu$m thick GEMs \cite{tamagawa2006}. The RIKEN-80T-LCP GEM is
   the finest pitch GEM with a thickness of 100 $\mu$m, and has become
   our standard for future space missions. For the rest of this paper,
   we mainly focus on this GEM.

  \subsection{Gain instability}
  
  It has been shown that the gain of GEMs produced by a chemical etching
  technique increases with time following the application of high
  voltage. Simon {\it et al.} stated that the gain of CERN and
  industrially produced Tech Etch GEMs in a triple GEM configuration
  increased by 30--80\% of the initial value over the 6 h following
  turning on of the high voltage \cite{simon2007}. A spacecraft in a low
  Earth orbit with an inclination angle of about 20--30$^{\circ}$, which
  is the standard orbit for scientific satellites launched from Japan or
  the United States, passes through the inner Van Allen belt known as
  the South Atlantic Anomaly (SAA) every 90 minutes, where a significant
  flux of charged particles (mainly protons) exist and interact in the
  detector. To avoid breakdown caused by the charged particles, gas
  detectors onboard satellites are usually turned off during passage
  through the SAA. In the operational mode where the applied voltage is
  turned on and off frequently, the change in gain related to turning on
  the high voltage is critical. It has been reported that the gain
  decrease of a charged-up detector when the voltage is switched off is
  very slow, of the order of many 10 s of hours
  \cite{altunbas2002}. However, we want to avoid those gain variations
  to reduce the systematic uncertainty of measurement.

  We have demonstrated that no significant gain increase is observed for
  the 50 $\mu$m thick laser etched GEMs under a very high rate of
  incident X-rays (10$^4$ counts mm$^{-2}$ sec$^{-1}$)
  \cite{tamagawa2006}. Here we demonstrated the gain stability for the
  thick-foil and fine-pitch LCP GEMs under a lower incident rate. Since
  the electron multiplication process in a gas detector is sensitive to
  the density of the gas, the measured gain has to be corrected by using
  the measured pressure and temperature. We have used the same
  correction function introduced in our previous study
  \cite{tamagawa2008}. The corrected effective gain ($G{\rm
  _{eff}^{corr}}$) normalized at a pressure of 760~Torr and a
  temperature of 300~K is given by;

  \begin{equation}
   G_{\rm eff}^{\rm corr}=\frac{G_{\rm eff}}{
    \exp\biggl(C\times\Bigl(\frac{1}{P/T}-\frac{1}{2.533}\Bigr)\biggr)}
  \end{equation}
  where $G_{\rm eff}$ is the measured effective gain introduced in
  \S\ref{sect:test_setup}, $C$ is a constant of 19.1~Torr K$^{-1}$ for
  the RIKEN-80T-LCP GEM in our setup, $P$ and $T$ are the pressure and
  temperature measured in Torr and K, respectively. During the
  experiment, the effective gain was around 10$^3$ and the GEM was
  irradiated with 5.9~keV X-rays at a rate of about 100~counts cm$^{-2}$
  sec$^{-1}$.

\begin{figure}[h]
   \begin{center}
    \includegraphics[width=1.0\textwidth]{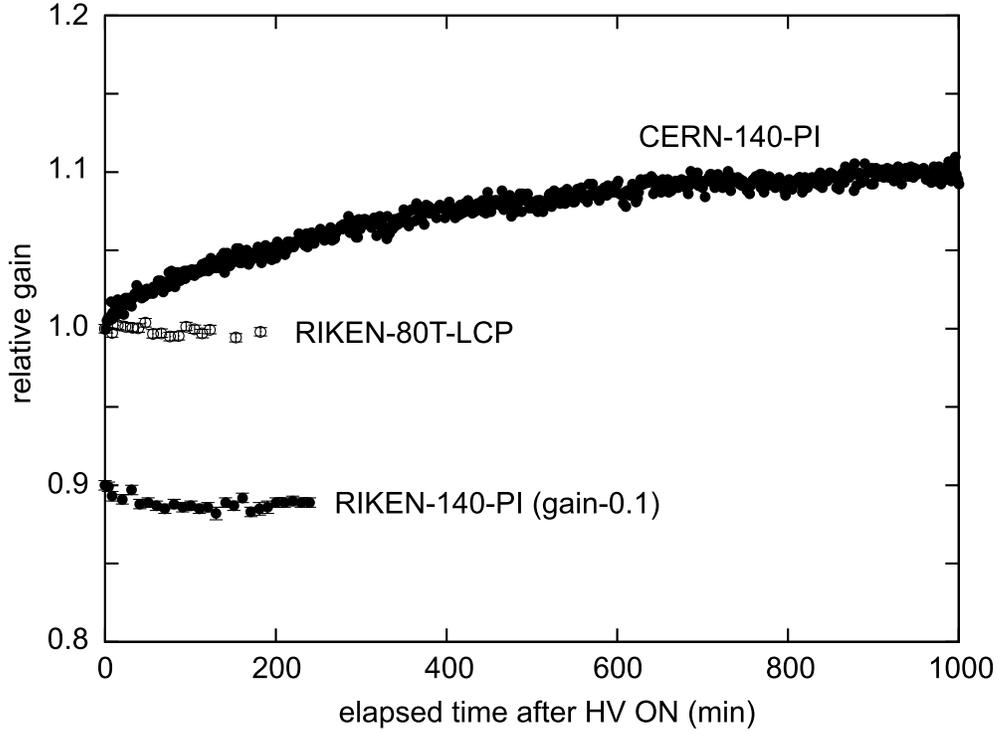}
   \end{center}
   \caption{Relative gain as a function of elapsed time after turning on
   of high voltage for RIKEN-80T-LCP and RIKEN-140-PI GEMs. The gain was
   normalized to 1 at the first measurement. For easier visibility, the
   gain of RIKEN-140-PI was offset by a value of $-0.1$. A correction
   for temperature and pressure was applied. The gain evolution of a
   CERN GEM (which had the same geometry as RIKEN-140-PI), as measured
   with our test setup, is shown in the figure. The effective gain of
   the measurements was around 10$^3$, and the count rate of irradiated
   5.9~keV X-rays was about 100~counts cm$^{-2}$ sec$^{-1}$.}
   \label{fig:gain_shortterm}
  \end{figure}

  The corrected gain of a single RIKEN-80T-LCP GEM, normalized to
  760~Torr and 300~K, as a function of the elapsed time after high
  voltage was applied, is shown in Figure~\ref{fig:gain_shortterm}. No
  time evolution was observed, and the standard deviation of the gain
  instability was 0.4\%. For comparison, the time evolution of the gain
  of the standard CERN GEM, which was the same geometry as RIKEN-140-PI,
  is superposed on the figure. The gain measurement was carried out
  under the same condition as for RIKEN-80T-LCP.

  As Benlloch {\it et al.} have pointed out \cite{benlloch1998}, a
  cylindrical hole shape formed by laser drilling is likely to suppress
  the charging-up of the surface of the side wall in the hole. The
  extremely good gain stability makes the laser etched GEMs optimal for
  applications where the gain stability is critical, such as for space
  applications.

  To demonstrate the gain stability of the PI GEMs, we have measured the
  time evolution of the gain for a single RIKEN-140-PI GEM, obtaining a
  standard deviation of 0.5\%. The measured gain is plotted in
  Figure~\ref{fig:gain_shortterm}. The result implies that the good gain
  stability results not from employing the new insulator material LCP,
  but from the cylindrical hole shape formed by the laser.

  \subsection{Gain uniformity}
  
\begin{figure}[h]
   \begin{center}
    \includegraphics[width=1.0\textwidth]{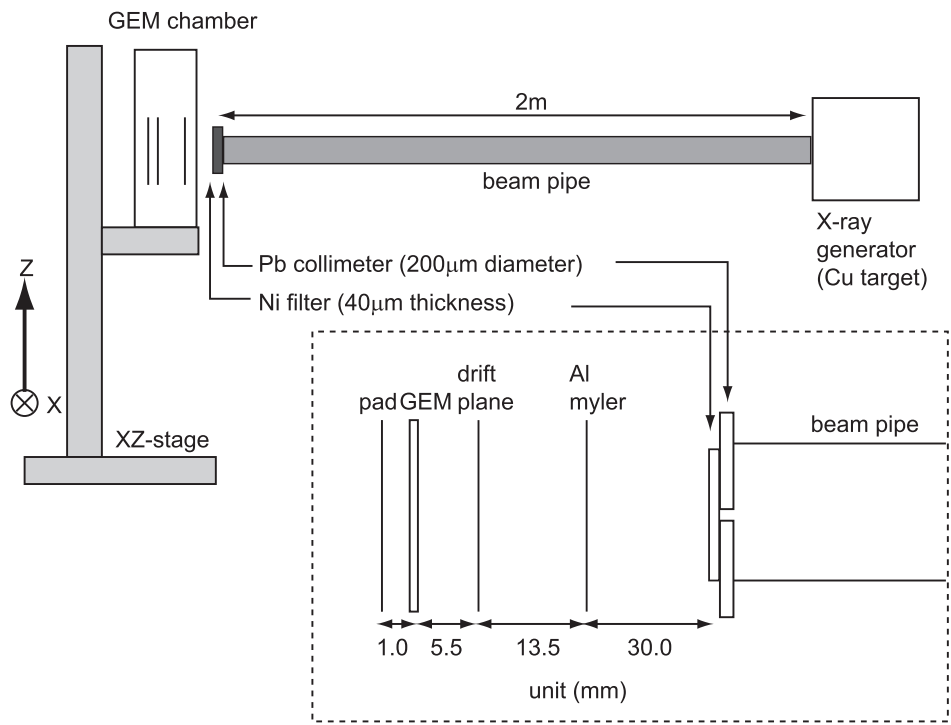}
   \end{center}
   \caption{A schematic view of the experimental setup for the gain
   uniformity mapping. An enlarged view around the end of the beam pipe
   is shown in the box surrounded by a dashed line.}
   \label{fig:exp_setup_uniformity}
  \end{figure}

  We have measured gain uniformity across the RIKEN-80T-LCP GEM by using
  a finely collimated X-ray beam. A schematic view of the experimental
  setup is shown in Figure~\ref{fig:exp_setup_uniformity}. The X-ray
  beam from an X-ray generator equipped with a copper target was
  connected to the experimental area with a 2 m long beam pipe. The beam
  size was controlled to a diameter of 200~$\mu$m by a pin-hole lead
  collimator attached to the beam pipe exit. Then, the beam was filtered
  by nickel foils to obtain a quasi-monochromatic beam with an energy of
  8~keV from Cu K$_{\alpha}$. The GEM chamber, which was identical to
  the one shown in Figure~\ref{fig:exp_setup}, was mounted on an
  XZ-stage to move horizontally and vertically, perpendicular to the
  collimated X-ray beam.
  
\begin{figure}[h]
   \begin{center}
    \includegraphics[width=0.7\textwidth]{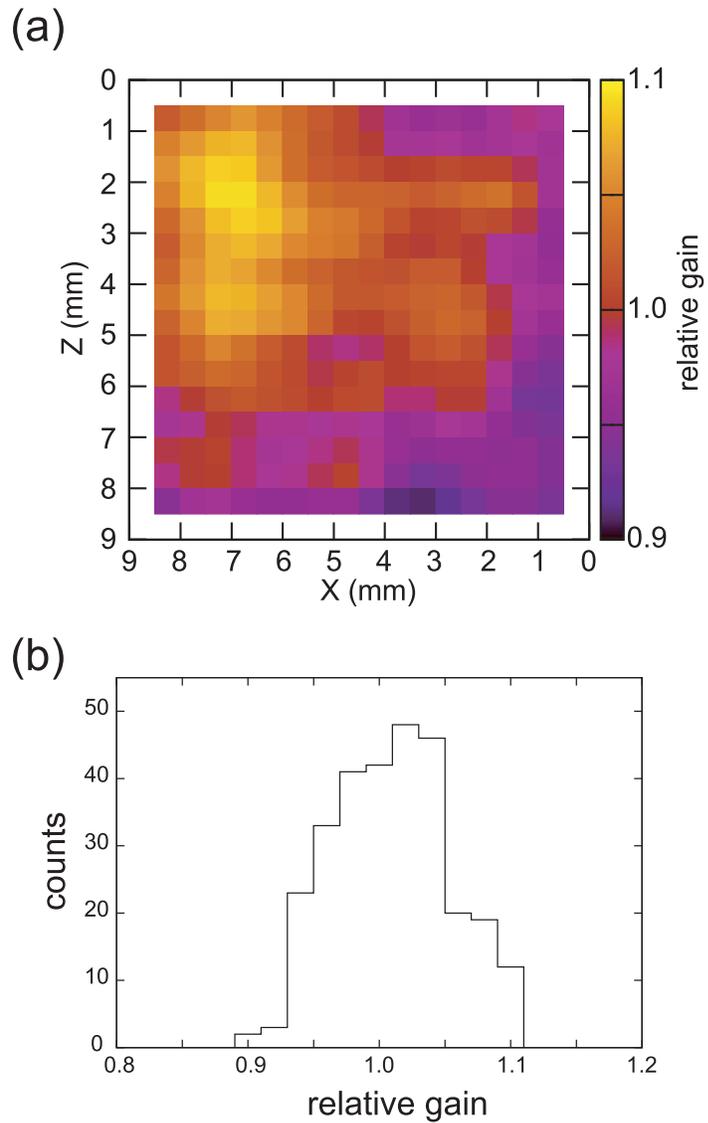}
   \end{center}
   \caption{(a) Map of the relative gain as a function of spatial
   location for a single RIKEN-80T-LCP GEM. The 9$\times$9~mm$^2$ region
   of the map is the same as shown in
   Figure~\ref{fig:hole_diameters}. (b) Distribution of the relative
   gain across the GEM. The standard deviation of the distribution was
   4.4\%.}  \label{fig:gain_uniformity}
  \end{figure}

  We have measured the gain within the central readout pad (9$\times$9
  mm$^2$) in 500~$\mu$m steps, i.e. we have measured the gain at
  17$\times$17 points within the pad. The gain was measured with an
  applied high voltage of 620~V between the electrodes of the GEM
  (gain$\sim$1000). A map of the measured gain is shown in Figure
  \ref{fig:gain_uniformity}a. A fractionally higher gain was observed at
  the upper-left region, while a slightly lower gain was observed at the
  lower-right region of the GEM. The gain was gradually changed across
  the GEM with a scale of a few millimeters. A histogram of the measured
  gain is shown in Figure \ref{fig:gain_uniformity}b with a standard
  deviation of 4.4\%. To check the systematic error, we have measured
  the gain map after rotating the GEM by 90 degrees relative to the
  chamber, obtaining the exact same map as obtained above. The gain maps
  were measured for several GEMs, showing standard deviations of 4--5\%.

\section{Distribution of hole diameters}
  
  We have measured the distribution of hole diameters across the
  RIKEN-80T-LCP GEM, that was used for the gain uniformity measurement
  above, with an automatic laser confocal microscope at the High Energy
  Accelerator Research Organization (KEK)
  \cite{yasuda2005,tawara2008}. The microscope took images with a linear
  image sensor, and then the images were analyzed with a software that
  measured hole diameters by matching an ellipse to each hole.

\begin{figure}[h]
   \begin{center}
   \includegraphics[width=0.7\textwidth]{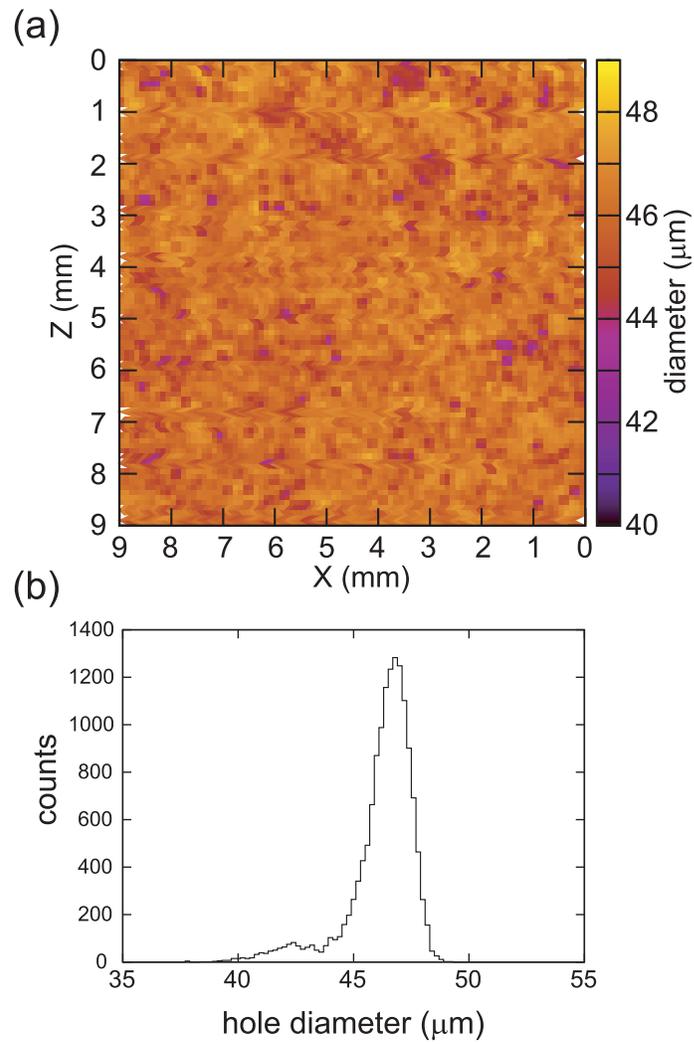}
   \end{center}
   \caption{(a) Spatial homogeneity of the hole diameter of
   RIKEN-80T-LCP. The color scale indicated on the right shows the hole
   diameter in units of $\mu$m. (b) The distribution of the hole
   diameters. The standard deviation was 3\%.}
   \label{fig:hole_diameters}
  \end{figure}

  Figure~\ref{fig:hole_diameters}a shows a contour map of the hole
  diameter distribution for the same region of the GEM shown in
  Figure~\ref{fig:gain_uniformity}a. The hole diameter distribution
  seems almost random across the GEM. The randomness indicates that the
  variation of the hole diameters is probably induced by the production
  accuracy. From Figures~\ref{fig:gain_uniformity}a and
  \ref{fig:hole_diameters}a, there was no obvious correlation between
  the gain and hole diameter distributions.

  The histogram of the distribution is shown in
  Figure~\ref{fig:hole_diameters}b. The mean value was 47~$\mu$m and the
  standard deviation was 3\%. Since we have measured the hole diameters
  on the surface of a copper electrode, the diameter was slightly larger
  than our design value (40~$\mu$m). The hole shape of the thin copper
  layer was conical with a surface diameter of 47~$\mu$m and a bottom
  (i.e. the boundary between the copper electrode and the LCP insulator)
  diameter of 38~$\mu$m.

\begin{figure}[h]
   \begin{center}
   \includegraphics[width=0.5\textwidth]{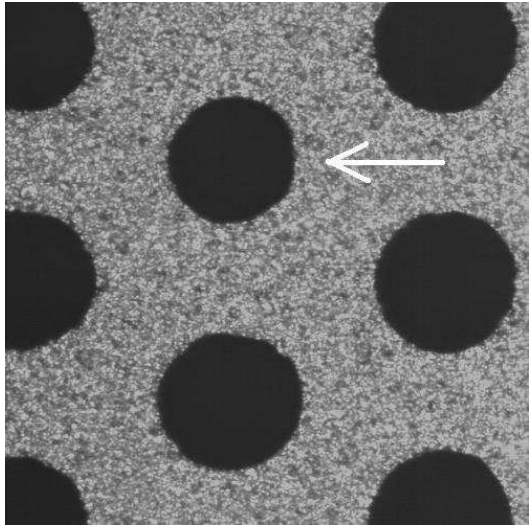}
   \end{center}
   \caption{A defect of a hole. The diameter of the hole indicated by
   the arrow is 42~$\mu$m, which is 5~$\mu$m smaller than average.}
   \label{fig:defects}
  \end{figure}

  A notable feature of the histogram is the second peak around
  43~$\mu$m. This corresponds to small dark spots in
  Figure~\ref{fig:hole_diameters}a, which are randomly distributed all
  over the GEM. A close-up microscopic view of one of the regions is
  shown in Figure~\ref{fig:defects}. Only the hole identified by an
  arrow has a smaller diameter than those of the other holes around
  it. We have measured the distribution of hole diameters for the other
  GEMs which had a nominal hole diameter of 70 $\mu$m, obtaining no such
  defects for those GEMs. We suspect that the defects came from an
  initial etching process which removed copper to form small diameter
  (40~$\mu$m) holes on the copper layers, and they will be improved in
  the next production.

 \section{Summary}

\begin{enumerate}
 \item We have produced thick-foil GEMs by using a CO$_2$ laser etching
 technique. We employed an insulator layer of liquid crystal polymer
 (LCP) instead of polyimide. Production yield was significantly improved
 by employing LCP, and the gain properties remained unchanged. This
 implies that LCP is an ideal material for producing GEMs.
 \item We have produced 100 $\mu$m thick-foil GEMs with hole pitches of
 140~$\mu$m and 80~$\mu$m. The achieved gain of the thick-foil GEMs was
 10$^4$ at an applied voltage of 700~V per GEM. The gain was 60 times
 higher than that of a double 50 $\mu$m thick GEM stack which was
 operated at the same combined voltage as the thick-foil GEMs.
 \item We have measured the gain stability of the thick-foil and
 fine-pitch GEM with 80 $\mu$m pitch, 40 $\mu$m hole, and 100 $\mu$m
 thick. No gain increase or decrease was observed after the high voltage
 was applied.  This was a common feature of the laser etched GEMs,
 probably due to the cylindrical shape of the hole formed by the laser
 drilling.
 \item We have measured the gain uniformity across the thick-foil and
 fine-pitch GEM. The standard deviation of the gain distribution in a
 9$\times$9~mm$^2$ region was about 4\%, and the gain gradually changed
 across the GEM on a scale of a few millimeters.
 \item We have measured the hole diameter distribution across the
 GEM. The standard deviation of the distribution was about 3\%. The
 distribution of hole diameters across the GEM is homogeneous, with no
 clear correlation between the hole diameter and gain maps.
\end{enumerate}

 \section*{Acknowledgement}

This work was partially supported by Japan Society for the Promotion of
Science (JSPS), Grant-in-Aid for Young Scientists (A)
(No. 19684008). A.H.  was supported by JSPS Research Fellowships. We
thank Professor F. Sauli and Mr. R. de Oliveira at CERN for their
valuable comments on our GEM productions. The authors thank
Mr. K. Miyamoto for his help on measuring GEM hole diameters at KEK. We
thank Professor S. Uno at KEK, Dr. J. Hill at NASA/GSFC, and P. Gandi at
RIKEN for discussions and comments.

\newpage

 \begin{table}[h]
  \begin{center}
   \caption{List of GEM foils examined in this study. All listed GEMs
   are produced by CO$_2$ laser etching.}
   \begin{tabular}{cccccc}
    \hline
    \makebox{name}& \makebox{hole pitch}& 
    \makebox{hole dia.} & \makebox{thickness} & 
    \makebox{insulator} & \makebox{thickness of Cu}\\
                  & \makebox{($\mu$m)}& \makebox{($\mu$m)}& 
    \makebox{($\mu$m)} & & \makebox{($\mu$m)} \\
    \hline\hline
    RIKEN-140-PI  & 140 & 70 & 50 & polyimide & 5\\
    RIKEN-140-LCP  & 140 & 70 & 50 & LCP & 9\\
    RIKEN-140T-LCP  & 140 & 70 & 100 & LCP & 9\\
    RIKEN-80T-LCP   &  50 & 30 & 100 & LCP &9\\
    \hline
   \end{tabular}
  \label{tab:gem_list}
  \end{center}
 \end{table}

\newpage

 \begin{table}[h]
   \caption{Typical mechanical, thermal, and hygroscopic properties of
   LCP and polyimide provided by SciEnergy Co., Ltd.} 
   \label{tab:lcp_parameters}
  \begin{minipage}{8cm}
   \begin{center}
    \begin{tabular}{cccc}
     \hline
     & \makebox{LCP} & \makebox{polyimide} & \makebox{note}\\
     \hline\hline
     Tensile Strength (MPa) & 200 & 274 &\\
     Tensile Elongation (\%) & 40 & 57 &\\
     Tensile Modulus (MPa) & 2900 & 4606 &\\
     CTE (ppm/K) & 20 & 20 &\\
     Thermal Conductivity (W/m$\cdot$K) & 0.5 & 0.2 &\\
     Water Absorption (\%) & 0.04 & 3.2 & 24~hrs in water\\
    Moisture absorption (\%) & $\le$0.04 & 1.5 & 24~hrs in
     50\%RH at 25
     $^\circ$C\\
     CHE (ppm/\%RH) & 1 & 28 &\\
     \hline
    \end{tabular}
   \footnotetext{CTE: Coefficient of Thermal Expansion}
   \footnotetext{RH: relative humidity}
   \footnotetext{CHE: Coefficient of Hydroscopic Expansion}
   \end{center}
  \end{minipage}
 \end{table}

\newpage

\newpage

\newpage

\newpage

\newpage

\newpage

\newpage

\newpage

\newpage

\end{document}